\pdfoutput=1

\documentclass[12pt,a4paper]{article}

\usepackage{ifthen} 
\newboolean{pdflatex}
\setboolean{pdflatex}{true} 

\newboolean{articletitles}
\setboolean{articletitles}{true} 

\newboolean{uprightparticles}
\setboolean{uprightparticles}{false} 

\newboolean{inbibliography}
\setboolean{inbibliography}{false} 

\usepackage{microtype}
\usepackage{lineno}  
\usepackage{xspace} 
\usepackage{caption} 

\usepackage{graphicx}  
\usepackage{color}
\usepackage{colortbl}
\graphicspath{{./figs/}} 

\usepackage{amsmath} 
\usepackage{amssymb}
\usepackage{amsfonts}
\usepackage{upgreek} 

\newcommand*\patchAmsMathEnvironmentForLineno[1]{%
\expandafter\let\csname old#1\expandafter\endcsname\csname #1\endcsname
\expandafter\let\csname oldend#1\expandafter\endcsname\csname
end#1\endcsname
 \renewenvironment{#1}%
   {\linenomath\csname old#1\endcsname}%
   {\csname oldend#1\endcsname\endlinenomath}%
}
\newcommand*\patchBothAmsMathEnvironmentsForLineno[1]{%
  \patchAmsMathEnvironmentForLineno{#1}%
  \patchAmsMathEnvironmentForLineno{#1*}%
}
\AtBeginDocument{%
\patchBothAmsMathEnvironmentsForLineno{equation}%
\patchBothAmsMathEnvironmentsForLineno{align}%
\patchBothAmsMathEnvironmentsForLineno{flalign}%
\patchBothAmsMathEnvironmentsForLineno{alignat}%
\patchBothAmsMathEnvironmentsForLineno{gather}%
\patchBothAmsMathEnvironmentsForLineno{multline}%
\patchBothAmsMathEnvironmentsForLineno{eqnarray}%
}

\usepackage{hyperref}    
\usepackage[all]{hypcap} 

\def\lhcb {\mbox{LHCb}\xspace}

\def\MagUp {\mbox{\em Mag\kern -0.05em Up}\xspace}

\ifthenelse{\boolean{uprightparticles}}%
{

 \def\Pmu         {\ensuremath{\upmu}\xspace}                 
 \def\Pnu         {\ensuremath{\upnu}\xspace}                 
                  
 \def\Ppi         {\ensuremath{\uppi}\xspace}

 \def\PDelta      {\ensuremath{\Delta}\xspace}                 
 \def\PXi      {\ensuremath{\Xi}\xspace}                 
 \def\PLambda      {\ensuremath{\Lambda}\xspace}                 
 \def\PSigma      {\ensuremath{\Sigma}\xspace}                 
 \def\POmega      {\ensuremath{\Omega}\xspace}                 
 \def\PUpsilon      {\ensuremath{\Upsilon}\xspace}                 
 
 \def\PB      {\ensuremath{\mathrm{B}}\xspace}                 
                  
 \def\PD      {\ensuremath{\mathrm{D}}\xspace}

 \def\PK      {\ensuremath{\mathrm{K}}\xspace}

 \def\PW      {\ensuremath{\mathrm{W}}\xspace}

 \def\Pb      {\ensuremath{\mathrm{b}}\xspace}                 
 \def\Pc      {\ensuremath{\mathrm{c}}\xspace}                 
 \def\Pd      {\ensuremath{\mathrm{d}}\xspace}                 
 \def\Pe      {\ensuremath{\mathrm{e}}\xspace}

 \def\Pi      {\ensuremath{\mathrm{i}}\xspace}

 \def\Pu      {\ensuremath{\mathrm{u}}\xspace}

}
{

 \def\Pmu         {\ensuremath{\mu}\xspace}                 
 \def\Pnu         {\ensuremath{\nu}\xspace}                 
                  
 \def\Ppi         {\ensuremath{\pi}\xspace}

 \mathchardef\PDelta="7101
 \mathchardef\PXi="7104
 \mathchardef\PLambda="7103
 \mathchardef\PSigma="7106
 \mathchardef\POmega="710A
 \mathchardef\PUpsilon="7107
                  
 \def\PB      {\ensuremath{B}\xspace}                 
                  
 \def\PD      {\ensuremath{D}\xspace}

 \def\PK      {\ensuremath{K}\xspace}

 \def\PW      {\ensuremath{W}\xspace}

 \def\Pb      {\ensuremath{b}\xspace}                 
 \def\Pc      {\ensuremath{c}\xspace}                 
 \def\Pd      {\ensuremath{d}\xspace}                 
 \def\Pe      {\ensuremath{e}\xspace}

 \def\Pi      {\ensuremath{i}\xspace}

 \def\Pu      {\ensuremath{u}\xspace}

}

\makeatletter
\ifcase \@ptsize \relax
  \newcommand{\miniscule}{\@setfontsize\miniscule{4}{5}}
\or
  \newcommand{\miniscule}{\@setfontsize\miniscule{5}{6}}
\or
  \newcommand{\miniscule}{\@setfontsize\miniscule{5}{6}}
\fi
\makeatother

\DeclareRobustCommand{\optbar}[1]{\shortstack{{\miniscule (\rule[.5ex]{1.25em}{.18mm})}
  \\ [-.7ex] $#1$}}

\def\epem       {{\ensuremath{\Pe^+\Pe^-}}\xspace}

\def\mun        {{\ensuremath{\Pmu^-}}\xspace} 

\def\ellm       {{\ensuremath{\ell^-}}\xspace}

\def\neub       {{\ensuremath{\overline{\Pnu}}}\xspace}

\def\neumb      {{\ensuremath{\neub_\mu}}\xspace}

\def\neulb      {{\ensuremath{\neub_\ell}}\xspace}

\def\W      {{\ensuremath{\PW}}\xspace}

\def\uquark    {{\ensuremath{\Pu}}\xspace}

\def\dquark    {{\ensuremath{\Pd}}\xspace}
\def\dquarkbar {{\ensuremath{\overline \dquark}}\xspace}

\def\cquark    {{\ensuremath{\Pc}}\xspace}

\def\bquark    {{\ensuremath{\Pb}}\xspace}
\def\bquarkbar {{\ensuremath{\overline \bquark}}\xspace}

\def\pion   {{\ensuremath{\Ppi}}\xspace}
\def\piz    {{\ensuremath{\pion^0}}\xspace}

\def\pip    {{\ensuremath{\pion^+}}\xspace}
\def\pim    {{\ensuremath{\pion^-}}\xspace}

\def\kaon    {{\ensuremath{\PK}}\xspace}
  \def\Kbar    {{\kern 0.2em\overline{\kern -0.2em \PK}{}}\xspace}

\def\KorKbar    {\kern 0.18em\optbar{\kern -0.18em K}{}\xspace}

\def\Kp      {{\ensuremath{\kaon^+}}\xspace}
\def\Km      {{\ensuremath{\kaon^-}}\xspace}

\def\KS      {{\ensuremath{\kaon^0_{\rm\scriptscriptstyle S}}}\xspace}

  \def\Dbar    {{\kern 0.2em\overline{\kern -0.2em \PD}{}}\xspace}
\def\D       {{\ensuremath{\PD}}\xspace}

\def\DorDbar    {\kern 0.18em\optbar{\kern -0.18em D}{}\xspace}
\def\Dz      {{\ensuremath{\D^0}}\xspace}

\def\B       {{\ensuremath{\PB}}\xspace}
\def\Bbar    {{\ensuremath{\kern 0.18em\overline{\kern -0.18em \PB}{}}}\xspace}

\def\BorBbar    {\kern 0.18em\optbar{\kern -0.18em B}{}\xspace}

\def\Bzb     {{\ensuremath{\Bbar{}^0}}\xspace}

\def\Bub     {{\ensuremath{\B^-}}\xspace}

\def\Bm      {{\ensuremath{\Bub}}\xspace}

  \def\Y#1S{\ensuremath{\PUpsilon{(#1S)}}\xspace}

\def\Lz          {{\ensuremath{\PLambda}}\xspace}
\def\Lbar        {{\ensuremath{\kern 0.1em\overline{\kern -0.1em\PLambda}}}\xspace}
\def\LorLbar    {\kern 0.18em\optbar{\kern -0.18em \PLambda}{}\xspace}

\def\Lb      {{\ensuremath{\Lz^0_\bquark}}\xspace}

\def\Lc      {{\ensuremath{\Lz^+_\cquark}}\xspace}

\newcommand{\decay}[2]{\ensuremath{#1\!\to #2}\xspace}         

\def\to                 {\ensuremath{\rightarrow}\xspace}

\def\Vub  {{\ensuremath{V_{\uquark\bquark}}}\xspace}

\def\AT#1     {\ensuremath{A_{\mathrm{T}}^{#1}}\xspace}           

\def\C#1      {\ensuremath{\mathcal{C}_{#1}}\xspace}                       
\def\Cp#1     {\ensuremath{\mathcal{C}_{#1}^{'}}\xspace}                    
\def\Ceff#1   {\ensuremath{\mathcal{C}_{#1}^{\mathrm{(eff)}}}\xspace}        
\def\Cpeff#1  {\ensuremath{\mathcal{C}_{#1}^{'\mathrm{(eff)}}}\xspace}       
\def\Ope#1    {\ensuremath{\mathcal{O}_{#1}}\xspace}                       
\def\Opep#1   {\ensuremath{\mathcal{O}_{#1}^{'}}\xspace}                    

\newcommand{\tev}{\ifthenelse{\boolean{inbibliography}}{\ensuremath{~T\kern -0.05em eV}\xspace}{\ensuremath{\mathrm{\,Te\kern -0.1em V}}}\xspace}
\newcommand{\gev}{\ensuremath{\mathrm{\,Ge\kern -0.1em V}}\xspace}
\newcommand{\mev}{\ensuremath{\mathrm{\,Me\kern -0.1em V}}\xspace}
\newcommand{\kev}{\ensuremath{\mathrm{\,ke\kern -0.1em V}}\xspace}
\newcommand{\ev}{\ensuremath{\mathrm{\,e\kern -0.1em V}}\xspace}
\newcommand{\gevc}{\ensuremath{{\mathrm{\,Ge\kern -0.1em V\!/}c}}\xspace}
\newcommand{\mevc}{\ensuremath{{\mathrm{\,Me\kern -0.1em V\!/}c}}\xspace}
\newcommand{\gevcc}{\ensuremath{{\mathrm{\,Ge\kern -0.1em V\!/}c^2}}\xspace}
\newcommand{\gevgevcccc}{\ensuremath{{\mathrm{\,Ge\kern -0.1em V^2\!/}c^4}}\xspace}
\newcommand{\mevcc}{\ensuremath{{\mathrm{\,Me\kern -0.1em V\!/}c^2}}\xspace}

\def\invfb   {\ensuremath{\mbox{\,fb}^{-1}}\xspace}

\def\mhz  {\ensuremath{{\rm \,MHz}}\xspace}
\def\khz  {\ensuremath{{\rm \,kHz}}\xspace}

\def\gsim{{~\raise.15em\hbox{$>$}\kern-.85em
          \lower.35em\hbox{$\sim$}~}\xspace}
\def\lsim{{~\raise.15em\hbox{$<$}\kern-.85em
          \lower.35em\hbox{$\sim$}~}\xspace}

\def\tell1  {TELL1\xspace}
\def\ukl1   {UKL1\xspace}

\usepackage{cite} 
\usepackage{mciteplus}

\def\vub {\ensuremath{|V_{\uquark\bquark}|}\xspace}
\def\vcb {\ensuremath{|V_{\cquark\bquark}|}\xspace}
\def\qq {\ensuremath{q^{2}}\xspace}
\def\PMuNu{\decay{\Lb}{p \mun \neumb}}
\def\LcMuNu{\decay{\Lb}{\Lc \mun \neumb}}

\def\PKpi{\decay{\Lb}{(\Lc \to p \Km \pip) \mun \neumb}}

\def\mcorr {\ensuremath{m_{\rm corr}}\xspace}

\begin{document}

\renewcommand{\thefootnote}{\fnsymbol{footnote}}
\setcounter{footnote}{1}

\begin{titlepage}
\pagenumbering{roman}

\vspace*{-1.5cm}
\centerline{\large EUROPEAN ORGANIZATION FOR NUCLEAR RESEARCH (CERN)}
\vspace*{1.5cm}
\hspace*{-0.5cm}
\begin{tabular*}{\linewidth}{lc@{\extracolsep{\fill}}r}
\ifthenelse{\boolean{pdflatex}}
{\vspace*{-2.7cm}\mbox{\!\!\!\includegraphics[width=.14\textwidth]{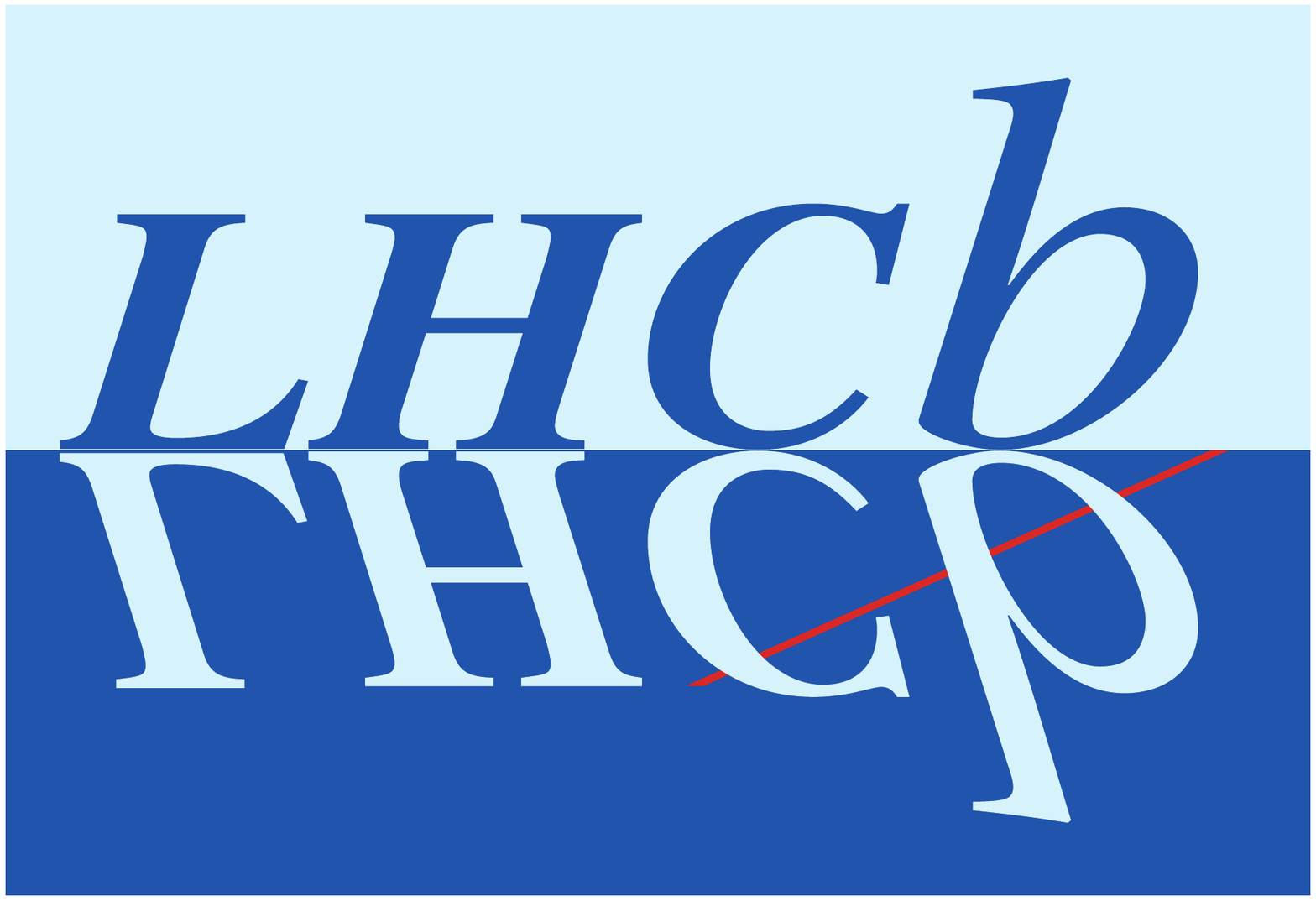}} & &}%
{\vspace*{-1.2cm}\mbox{\!\!\!\includegraphics[width=.12\textwidth]{lhcb-logo.eps}} & &}%
\\
 & & CERN-PH-EP-2015-084 \\  
 & & LHCb-PAPER-2015-013 \\  
 & & 27 July, 2015 \\ 
 & & \\
\end{tabular*}

\vspace*{3.0cm}

{\bf\boldmath\huge
\begin{center}
Determination of the quark coupling strength $|V_{ub}|$ using baryonic decays
\end{center}
}

\vspace*{1.8cm}

\begin{center}
The LHCb collaboration\footnote{Authors are listed at the end of this paper.}
\end{center}

\vspace{\fill}

\begin{abstract}
  \noindent
  In the Standard Model of particle physics, the strength of the couplings of the $b$ quark to the $u$ and $c$ quarks,
  $|V_{ub}|$ and $|V_{cb}|$, are governed by the coupling of the quarks to the Higgs boson. Using
  data from the LHCb experiment at the Large Hadron Collider,
   the probability for the $\Lambda^0_b$ baryon to decay into the $p \mu^-
  \overline{\nu}_\mu$ final state relative to the $\Lambda^+_c \mu^-
  \overline{\nu}_\mu$ final state is measured. Combined with theoretical calculations 
  of the strong interaction and a previously measured value of $|V_{cb}|$, the first 
  $|V_{ub}|$ measurement to use a baryonic decay is performed. This measurement is consistent with previous determinations of $|V_{ub}|$ 
  using $B$ meson decays to specific final states and confirms the existing incompatibility 
  with those using an inclusive sample of final states. 
\end{abstract}
\vspace*{1.8cm}

\begin{center}
  \href{http://dx.doi.org/10.1038/nphys3415}{Published in Nature Physics}
\end{center}

\vspace{\fill}

{\footnotesize 
\centerline{\copyright~CERN on behalf of the \lhcb collaboration, licence \href{http://creativecommons.org/licenses/by/4.0/}{CC-BY-4.0}.}}
\vspace*{2mm}

\end{titlepage}


\newpage
\setcounter{page}{2}
\mbox{~}

\cleardoublepage

\twocolumn

\renewcommand{\thefootnote}{\arabic{footnote}}
\setcounter{footnote}{0}



\pagestyle{plain} 
\setcounter{page}{1}
\pagenumbering{arabic}


%

\noindent In the Standard Model~(SM) of particle physics, the decay of one quark to another by the emission of a virtual \W boson is described by the $3 \times 3$ unitary Cabibbo-Kobayashi-Maskawa (CKM) matrix~\cite{Cabibbo,KM}. This matrix arises from the coupling of the quarks to the Higgs boson. While the SM does not predict the values of the four free parameters of the CKM matrix, the measurements of these parameters in different processes should be consistent with each other. If they are not, it is a sign of physics beyond the SM.  In global fits combining all available measurements~\cite{Charles:2015gya, UTFit}, the sensitivity of the overall consistency check is limited by the precision in the measurements of the magnitude and phase of the matrix element \Vub, which describes the transition of a \bquark quark to a \uquark quark.

The magnitude of \Vub can be measured via the semileptonic quark-level transition \mbox{$\bquark\to\uquark\ellm\neulb$}. Semileptonic decays are used to minimise the uncertainties arising from the interaction of the strong force, described by quantum chromodynamics (QCD), between the final-state quarks. For the measurement of the magnitude of \Vub, as opposed to measurements of the phase, all decays of the \bquark quark, and the equivalent \bquarkbar quark, can be considered together. There are two complementary methods to perform the measurement. From an experimental point of view, the simplest is to measure the branching fraction (probability to decay to a given final state) of a specific (exclusive) decay. An example is the decay of a \Bzb (\bquark\dquarkbar) meson to the final state $\pip\ellm\neub$, where the influence of the strong interaction on the decay, encompassed by a $\Bzb\to\pip$ form factor, is predicted by non-perturbative techniques such as lattice QCD (LQCD)~\cite{Lattice} or QCD sum rules~\cite{SumRules}. The world average from Ref.~\cite{HFAG} for this method, using the decays $\Bzb\to\pip\ellm\neub$ and $\Bm\to\piz\ellm\neub$, is $\vub = (3.28\pm0.29)\times10^{-3}$, where the most precise experimental inputs come from the BaBar~\cite{Lees:2012vv,delAmoSanchez:2010af} and Belle~\cite{Ha:2010rf,Sibidanov:2013rkk} experiments. The uncertainty is dominated by the LQCD calculations, which have recently been updated~\cite{Flynn:2015mha,Lattice:2015tia} and result in larger values of \Vub than the average given in Ref.~\cite{HFAG}. The alternative method is to measure the differential decay rate in an inclusive way over all possible \B meson decays containing the $\bquark\to\uquark\ellm\neub$ quark level transition. This results in $\vub = (4.41\pm0.15^{+0.15}_{-0.17})\times10^{-3}$~\cite{PDG2014}, where the first uncertainty arises from the experimental measurement and the second from theoretical calculations. The discrepancy between the exclusive and inclusive \vub determinations is approximately three standard deviations and has been a long-standing puzzle in flavour physics. Several explanations have been proposed, such as the presence of a right-handed (vector plus axial-vector) coupling as an extension of the SM beyond the left-handed (vector minus axial-vector) \W coupling~\cite{Chen:2008se,PhysRevD.81.031301,Buras:2010pz,Crivellin:2014zpa}. A similar discrepancy also exists between exclusive and inclusive measurements of $\vcb$ (the coupling of the \bquark quark to the \cquark quark)~\cite{PDG2014}.

This article describes a measurement of the ratio of branching fractions of the \Lb (\bquark\uquark\dquark) baryon into the $p\ellm\neub$ and $\Lc\ellm\neub$ final states. This is performed using proton-proton collision data from the \lhcb detector, corresponding to 2.0\invfb of integrated luminosity collected at a centre-of-mass energy of $8\tev$. The $\bquark\to\uquark$ transition, \PMuNu, has not been considered before as \Lb baryons are not produced at an \epem\ \B-factory; however, at the LHC, they constitute around 20\% of the $b$-hadrons produced~\cite{LHCB-PAPER-2011-018}. These measurements together with recent LQCD calculations~\cite{LQCDdyn} allow for the determination of $\vub^2/\vcb^2$ according to
\begin{equation}
\label{eq:one}
\frac{\vub^2}{\vcb^2} =  \frac{\mathcal{B}(\PMuNu)}{\mathcal{B}(\LcMuNu)} R_{\rm FF}
\end{equation}
where $\mathcal{B}$ denotes the branching fraction and $R_{\rm FF}$ is a ratio of the relevant form factors, calculated using LQCD. 
This is then converted into a measurement of $\vub$ using the existing measurements of 
$\vcb$ obtained from exclusive decays. The normalisation to the \LcMuNu decay cancels many experimental uncertainties, including the uncertainty on the total production rate of \Lb baryons. At the LHC, the number of signal candidates is large, allowing the optimisation of the event selection and the analysis approach to minimise systematic effects.

The \lhcb detector~\cite{Alves:2008zz,LHCb-DP-2014-002} is one of the four major detectors at the Large Hadron Collider. It is instrumented in a cone around the proton beam axis, covering the angles between 10 and 250~mrad, where most \bquark-hadron decays produced in proton-proton collisions occur. The detector includes a high-precision tracking system with a dipole magnet, providing a measurement of momentum and impact parameter (IP), defined for charged particles as the minimum distance of a track to a primary proton-proton interaction vertex (PV). Different types of charged particles are distinguished using information from two ring-imaging Cherenkov detectors, a calorimeter and a muon system. Simulated samples of specific signal and background decay modes of \bquark hadrons are used at many stages throughout the analysis. These simulated events model the experimental conditions in full detail, including the proton-proton collision, the decay of the particles, and the response of the detector. The software used is described in Refs.~\cite{Sjostrand:2006za, LHCb-PROC-2010-056, Lange:2001uf, Golonka:2005pn, Allison:2006ve, Agostinelli:2002hh, LHCb-PROC-2011-006}.

Candidates of the signal modes are required to pass a trigger system~\cite{LHCb-DP-2012-004} which reduces in real-time the rate of recorded collisions (events) from the 40\mhz read-out clock of the LHC to around 4\khz. For this analysis, the trigger requires a muon with a large momentum transverse to the beam axis that at the same time forms a good vertex with another track in the event. This vertex should be displaced from the PVs in the event. The identification efficiency for these high momentum muons is 98\%.
  
In the selection of the final states, stringent particle identification (PID) requirements are applied to the proton. These criteria are accompanied by a requirement that its momentum is greater than $15 \gevc$ as the PID performance is most effective for protons above the momentum threshold to produce Cherenkov light. The $p\mun$ vertex fit is required to be of good quality, which reduces background from most of the $\bquark\to\cquark\mun\neumb$ decays as the resulting ground state charmed hadrons have significant lifetime.

To reconstruct \PKpi candidates, two additional tracks, positively identified as a pion and kaon, are combined with the proton to form a $\Lc\to p \Km \pip$ candidate. These are reconstructed from the same $p\mun$ vertex as the \PMuNu signal to minimise systematic uncertainties. As the lifetime of the \Lc is short compared to other weakly decaying charm hadrons, the requirement has an acceptable efficiency.

\begin{figure}[tb]
  \centering
  \includegraphics[width=0.45\textwidth]{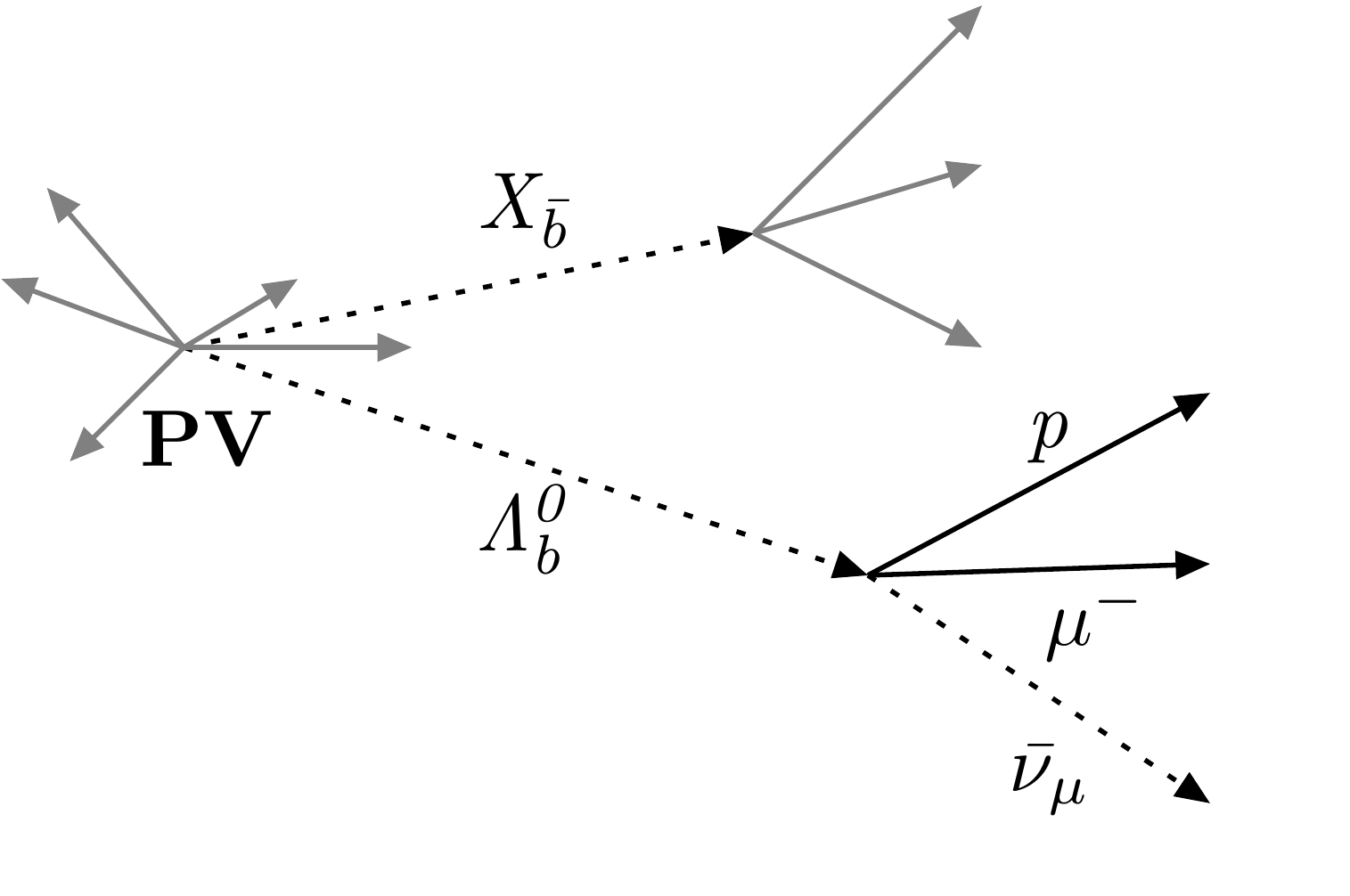} 
  \includegraphics[width=0.45\textwidth]{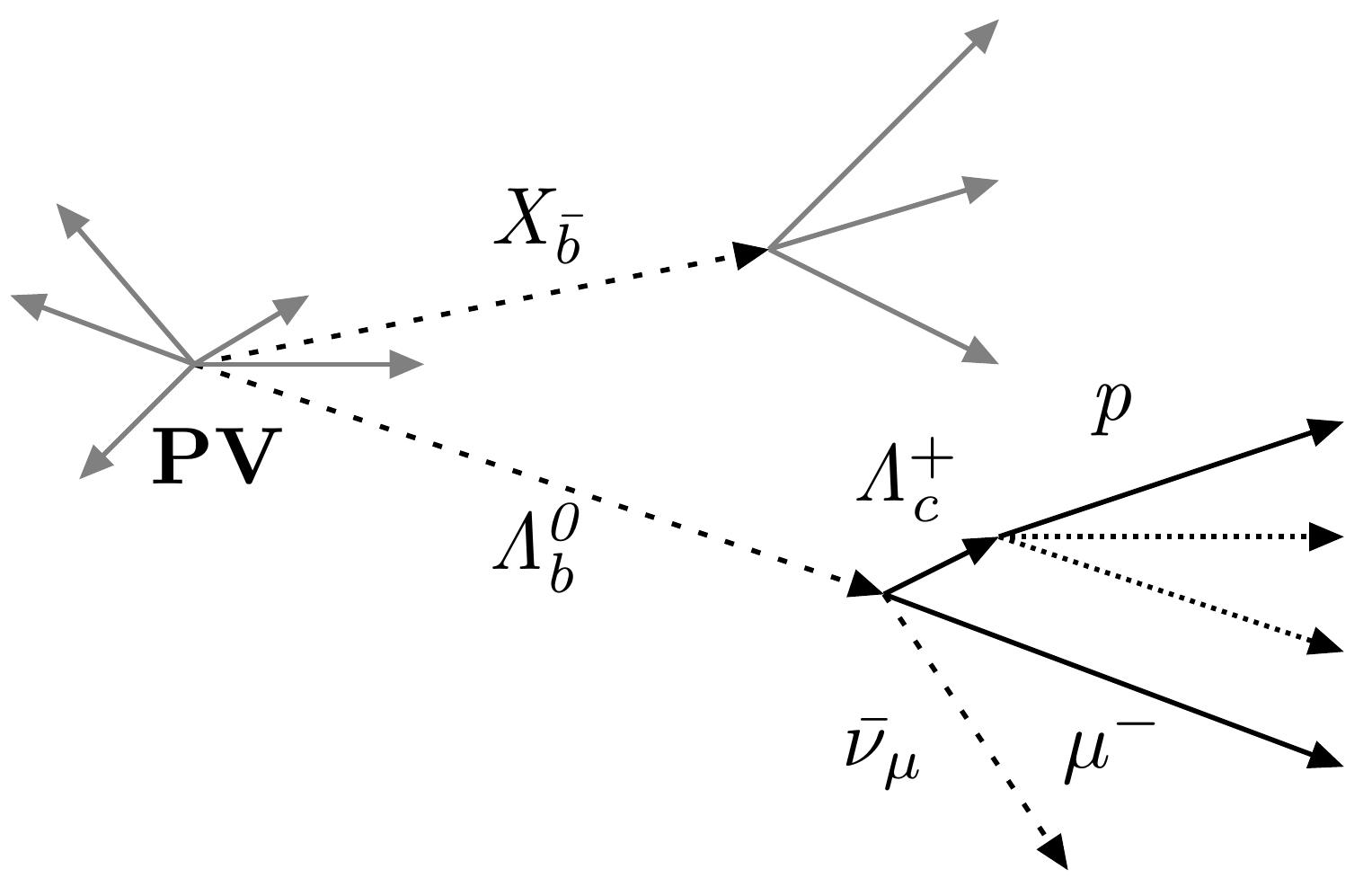} 
  \caption{\textbf{Diagram illustrating the topology for the (top) signal and (bottom) background decays.} The $\Lb$ baryon travels about 1\,cm on average before decaying; its flight direction is indicated in the diagram. In the \PMuNu signal case, the only other particles present are typically reconstructed far away from the signal, which are shown as grey arrows. For the background from \Lc decays, there are particles which are reconstructed in close proximity to the signal and which are indicated as dotted arrows.}
  \label{fig:iso}
\end{figure}

There is a large background from \bquark-hadron decays with additional charged tracks in the decay products, as illustrated in Fig.~\ref{fig:iso}. To reduce this background, a multivariate machine learning algorithm (a boosted decision tree, BDT~\cite{Breiman,AdaBoost}) is employed to determine the compatibility of each track from a charged particle in the event to originate from the same vertex as the signal candidate. This \emph{isolation BDT} includes variables such as the change in vertex quality if the track is combined with the signal vertex, as well as kinematic and IP information of the track that is tested. For the BDT, the training sample of well isolated tracks consists of all tracks apart from the signal decay products in a sample of simulated \PMuNu events. The training sample of non-isolated tracks consists of the tracks from charged particles in the decay products $X$ in a sample of simulated $\Lb\to(\Lc \to p X) \mun\neumb$ events. The BDT selection removes 90\% of background with additional charged particles from the signal vertex while it retains more than 80\% of signal. The same isolation requirement is placed on both the \PMuNu and \PKpi decay candidates, where the pion and kaon are ignored in the calculation of the BDT response for the \PKpi case.

The \Lb mass is reconstructed using the so-called \emph{corrected mass}~\cite{Abe:1997sb}, defined as
\begin{displaymath}
  \mcorr = \sqrt{ m_{h\mu}^{2} + p_{\rm{\perp}}^{2}} + p_{\rm{\perp}},
\end{displaymath}
where $m_{h\mu}$ is the visible mass of the ${h\mu}$ pair and $p_{\rm{\perp}}$ is the momentum of the $h\mu$ pair transverse to the \Lb flight direction, where $h$ represents either the proton or \Lc candidate.  The flight direction is measured using the PV and \Lb vertex positions. The uncertainties on the PV and the \Lb vertex are estimated for each candidate and propagated to the uncertainty on \mcorr; the dominant contribution is from the uncertainty in the \Lb vertex.

Candidates with an uncertainty of less than $100\mevcc$ on the corrected mass are selected for the \PMuNu decay. This selects only 23\% of the signal; however, the separation between signal and background for these candidates is significantly improved and the selection thus reduces the dependence on background modelling.
 
\begin{figure}[tb]
  \centering
  \includegraphics[width=0.45\textwidth]{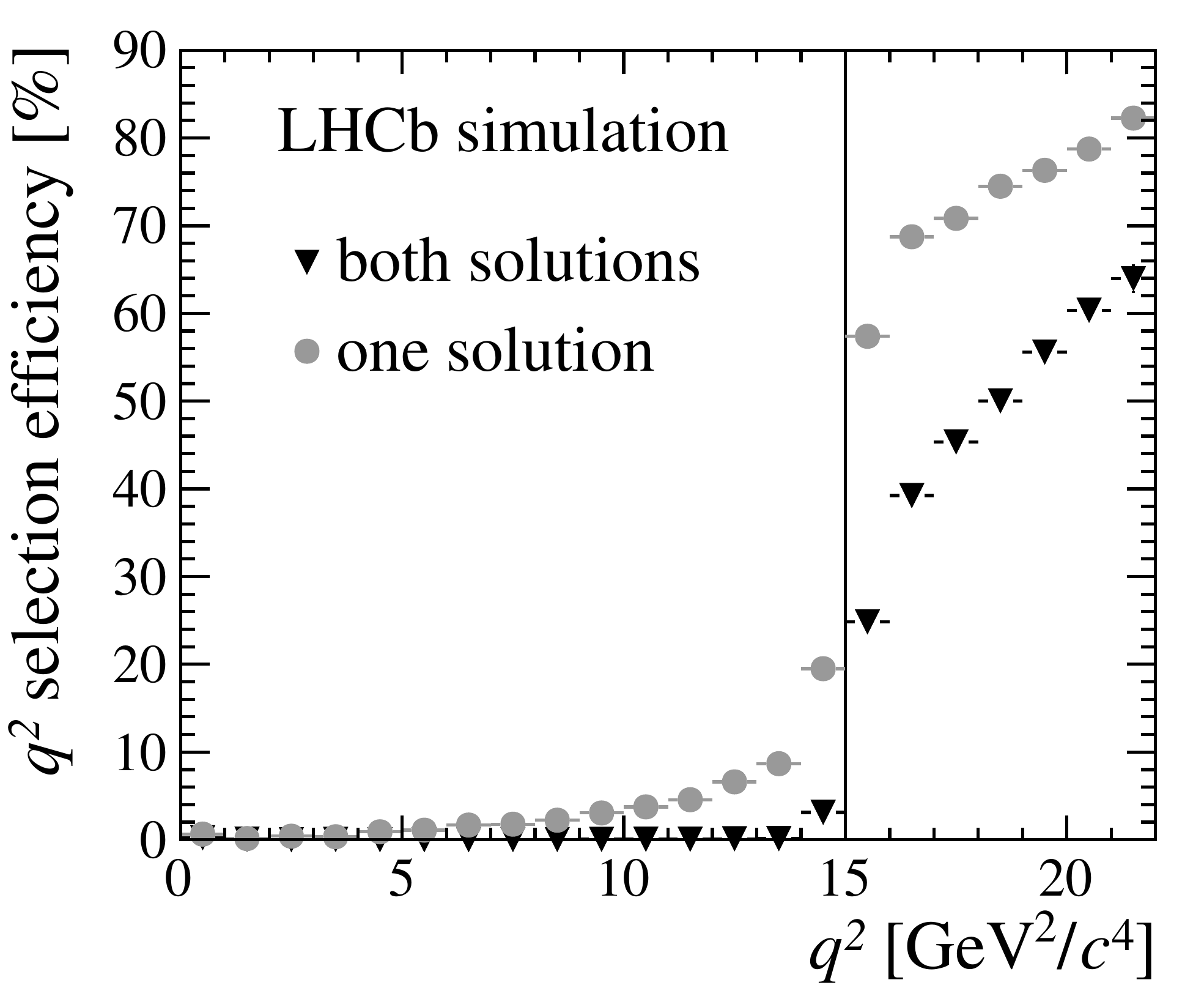} 
  \caption{\textbf{Illustrating the method used to reduce the number
      of selected events from the \boldmath{\qq} region where lattice
      QCD has high uncertainties.} The efficiency of simulated \PMuNu
    candidates as a function of \qq. For the case where one \qq
    solution is required to be above 15\gevgevcccc (marked by the
    vertical line), there is still significant efficiency for signal
    below this value, whereas, when both solutions have this
    requirement, only a small amount of signal below 15\gevgevcccc is
    selected.}
  \label{fig:migration}
\end{figure}

The LQCD form-factors that are required to calculate \vub are most precise in the kinematic region where \qq, the invariant mass squared of the muon and the neutrino in the decay, is high. The neutrino is not reconstructed, but \qq can still be determined using the \Lb flight direction and the \Lb mass, but only up to a two-fold ambiguity. The correct solution has a resolution of about 1\gevgevcccc, while the wrong solution has a resolution of 4\gevgevcccc. To avoid influence on the measurement by the large uncertainty in form factors at low \qq, both solutions are required to exceed 15\gevgevcccc for the \PMuNu decay and 7\gevgevcccc for the \PKpi decay. Simulation shows that only 2\% of \PMuNu decays and 5\% of \LcMuNu decays with \qq values below the cut values pass the selection requirements. The effect of this can be seen in Fig.~\ref{fig:migration}, where the efficiency for signal below 15\gevgevcccc is reduced significantly if requirements are applied on both solutions. It is also possible that both solutions are imaginary due to the limited detector resolution. Candidates of this type are rejected. The overall \qq selection has an efficiency of 38\% for \PMuNu and 39\% for \LcMuNu decays in their respective high-\qq regions.

\begin{figure}[tb]
  \centering
  \includegraphics[width=0.45\textwidth]{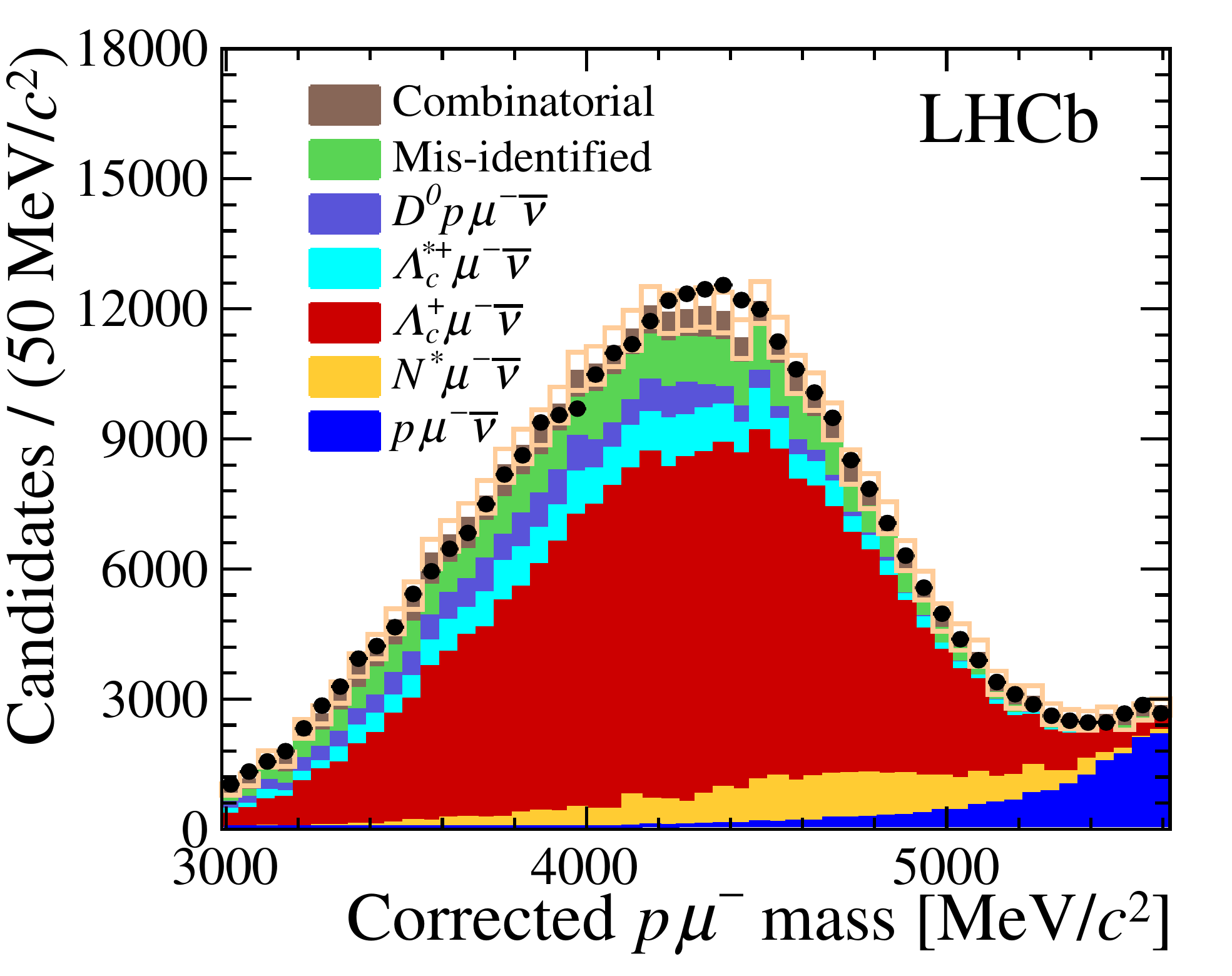} 
  \includegraphics[width=0.45\textwidth]{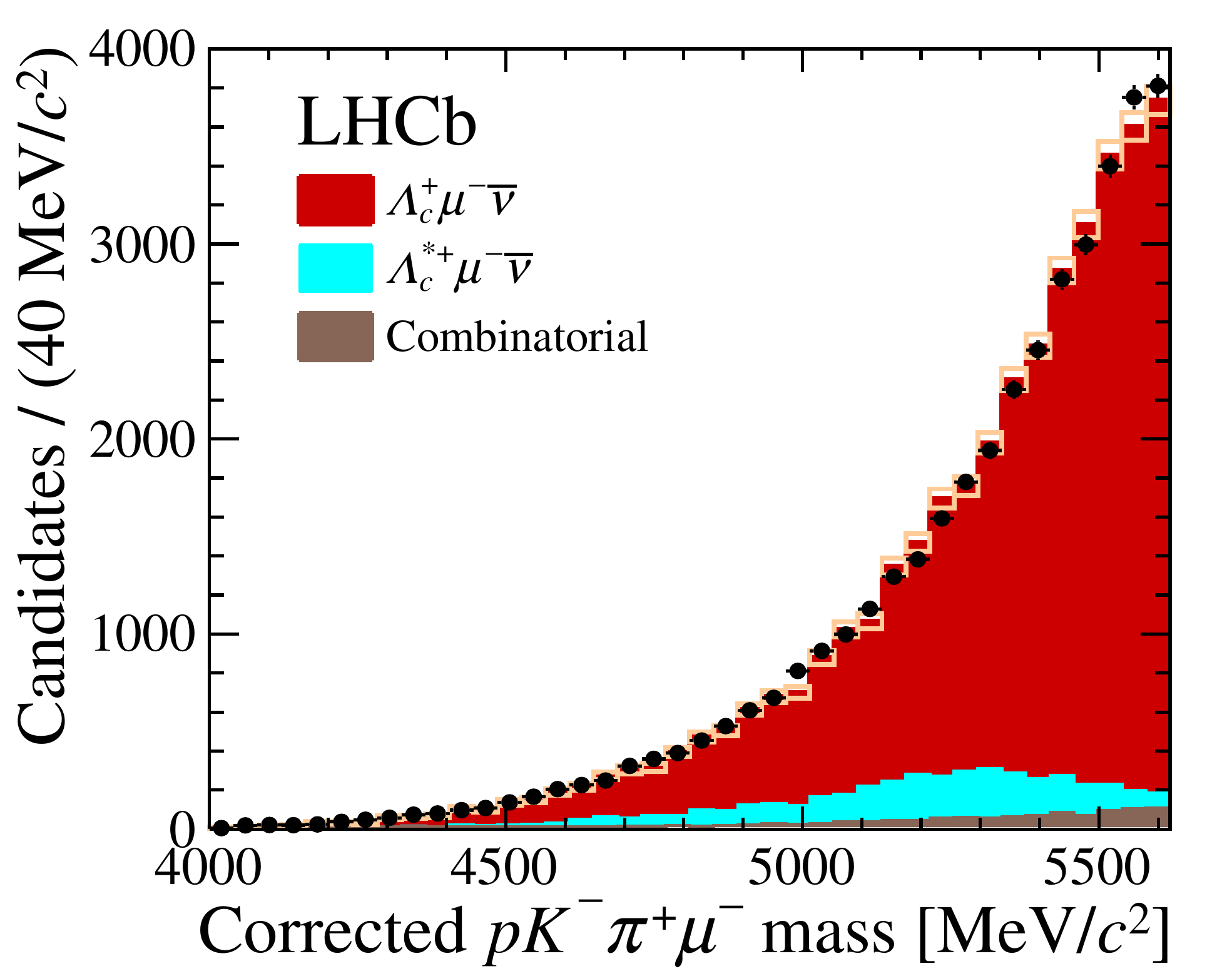} 
  \caption{\textbf{Corrected mass fit used for determining signal
      yields.} Fits are made to (top) \PMuNu and (bottom) \PKpi
    candidates. The statistical uncertainties arising from the finite
    size of the simulation samples used to model the mass shapes are
    indicated by open boxes while the data are represented by the
    black points. The statistical uncertainty on the data points is
    smaller than the marker size used. The different signal and
    background components appear in the same order in the fits and the
    legends. There are no data above the nominal \Lb mass due to the
    removal of unphysical \qq solutions.}
  \label{fig:fit}
\end{figure}
The mass distributions of the signal candidates for the two decays are shown in Fig.~\ref{fig:fit}. The signal yields are determined from separate $\chi^2$ fits to the \mcorr distributions of the \PMuNu and \PKpi candidates. The shapes of the signal and background components are modelled using simulation, where the uncertainties coming from the finite size of the simulated samples are propagated in the fits. The yields of all background components are allowed to vary within uncertainties obtained as described below.

For the fit to the \mcorr distribution of the \PMuNu candidates, many sources of background are accounted for. The largest of these is the cross-feed from \LcMuNu decays, where the \Lc decays into a proton and other particles that are not reconstructed. The amount of background arising from these decay modes is estimated by fully reconstructing two \Lc decays in the data. The background where the additional particles include charged particles originating directly from the \Lc decay is estimated by reconstructing \mbox{\PKpi} decays, whereas the background where only neutral particles come directly from the \Lc decay is estimated by reconstructing \mbox{$\Lb\to(\Lc\to p \KS)\mun\neumb$} decays. These two background categories are separated because the isolation BDT significantly reduces the charged component but has no effect on the neutral case. For the rest of the \Lc decay modes, the relative branching fraction between the decay and either the $\Lc\to p \Km \pip$ or $\Lc\to p \KS$ decay modes, as appropriate, is taken from Ref.~\cite{PDG2014}. For some neutral decay modes, where only the corresponding mode with charged decay particles is measured, assumptions based on isospin symmetry are used. In these decays, an uncertainty corresponding to 100\% of the branching fraction is allowed for in the fit. Background from $\Lb\to \Dz p\mun\neumb$ decays is constrained in a similar way to the \Lc charged decay modes, with the normalisation done relative to $\Lb\to \Dz(\to \Km \pip) p\mun\neumb$ decays reconstructed in the data.

Any background with a \Lc baryon may also arise from decays of the type $\Lb\to(\Lz^{*+}_{c}\to\Lc\pi\pi)\mun\neumb$, where $\Lz^{*+}_{c}$ represents the $\Lz_{c}(2595)^{+}$ or $\Lz_{c}(2625)^{+}$ resonances as well as non-resonant contributions. The proportions between the \mbox{$\Lb\to(\Lz^{*+}_{c}\to\Lc\pi\pi)\mun\neumb$} and the \mbox{\LcMuNu} backgrounds are determined from the fit to the \mbox{$\Lb\to(\Lc\to p\Km\pip)\mun\neumb$} \mcorr distribution and then used in the \PMuNu fit.

The decays $\Lb\to N^{*}\mun\neumb$, where the $N^{*}$ baryon decays into a proton and other non-reconstructed particles, are very similar to the signal decay and have poorly known branching fractions. The $N^{*}$ resonance represents any of the states $N(1440)$, $N(1520)$, $N(1535)$ or $N(1770)$. None of the $\Lb\to N^{*}\mun\neumb$ decays have been observed and the \mcorr shape of these decays is obtained using simulation samples generated according to the quark-model prediction of the form factors and branching fractions~\cite{SLBaryon}. A 100\% uncertainty is allowed for in the branching fractions of these decays.

Background where a pion or kaon is mis-identified as a proton originates from various sources and is measured by using a special data set where no PID is applied to the proton candidate. Finally, an estimate of combinatorial background, where the proton and muon originate from different decays, is obtained from a data set where the proton and muon have the same charge. The amount and shape of this background are in good agreement between the same-sign and opposite-sign $p\mu$ samples for corrected masses above 6\gevcc.

For the \PKpi yield, the reconstructed $p\Km\pip$ mass is studied to determine the level of combinatorial background. The \Lc signal shape is modelled using a Gaussian function with an asymmetric power-law tail, and the background is modelled as an exponential function. Within a selected signal region of $30\mevcc$ from the known \Lc mass the combinatorial background is 2\% of the signal yield.  Subsequently, a fit is performed to the \mcorr distribution for \PKpi candidates, as shown in Fig.~\ref{fig:fit}, which is used to discriminate between \LcMuNu and $\Lb\to(\Lz^{*+}_{c}\to\Lc\pi\pi)\mun \neumb$ decays.

The \PMuNu and \PKpi yields are 17,687$\pm733$ and 34,255$\pm571$, respectively. This is the first observation of the decay \PMuNu.

The \PMuNu branching fraction is measured relative to the \PKpi branching fraction. The relative efficiencies for reconstruction, trigger and final event selection are obtained from simulated events, with several corrections applied to improve the agreement between the data and the simulation. These correct for differences between data and simulation in the detector response and differences in the \Lb kinematic properties for the selected \PMuNu and \PKpi candidates. The ratio of efficiencies is $3.52\pm0.20$, with the sources of the uncertainty described below. 

Systematic uncertainties associated with the measurement are summarised in Table~\ref{tab:system}.  The largest uncertainty originates from the $\Lc\to p \Km \pip$ branching fraction, which is taken from Ref.~\cite{Zupanc:2013iki}. This is followed by the uncertainty on the trigger response, which is due to the statistical uncertainty of the calibration sample. Other contributions come from the tracking efficiency, which is due to possible differences between the data and simulation in the probability of interactions with the material of the detector for the kaon and pion in the \PKpi decay.  Another systematic uncertainty is assigned due to the limited knowledge of the momentum distribution for the $\Lc \to p \Km \pip$ decay products. Uncertainties related to the background composition are included in the statistical uncertainty for the signal yield through the use of nuisance parameters in the fit. The exception to this is the uncertainty on the $\Lb\to N^{*}\mun\neumb$ mass shapes due to the limited knowledge of the form factors and widths of each state, which is estimated by generating pseudoexperiments and assessing the impact on the signal yield.
\begin{table}[t]

\newcommand\Tstrut{\rule{0pt}{2.4ex}}
\newcommand\TTstrut{\rule{0pt}{3.2ex}}
\newcommand\Bstrut{\rule[-1.2ex]{0pt}{0pt}}
\newcommand\BBstrut{\rule[-1.8ex]{0pt}{0pt}}

\caption{\textbf{Summary of systematic uncertainties.} The table shows the relative systematic uncertainty on the ratio of the \PMuNu and \LcMuNu branching fractions broken into its individual contributions. The total is obtained by adding them in quadrature. Uncertainties on the background levels are not listed here as they are incorporated into the fits.}
      \label{tab:system}
      \resizebox{0.47\textwidth}{!} {
    \begin{tabular}{l c}
    \hline
      Source & \hspace{-1.1cm}Relative uncertainty (\%)   \\
      \hline
      \vspace{1mm}
      \hspace{-1mm}$\mathcal{B}(\Lc\to p \Kp \pim)$ & $\phantom{}^{+4.7}_{-5.3}\phantom{.}$ \Tstrut  \\
      Trigger & 3.2 \\
      Tracking & 3.0 \\
      \Lc selection efficiency & 3.0 \\
      $\Lb\to N^{*}\mun\neumb$ shapes & 2.3 \\
      \Lb lifetime & 1.5 \\
      Isolation & 1.4 \\
      Form factor & 1.0 \\
      \Lb kinematics & 0.5 \\
      \qq migration & 0.4 \\
      PID & 0.2 \\
      \hline
      Total & \hspace{-0.07cm}$\phantom{}^{+7.8}_{-8.2}\phantom{.}$\Tstrut \\
      \hline
 
    \end{tabular} }
\end{table}

Smaller uncertainties are assigned for the following effects: the uncertainty in the \Lb lifetime; differences in data and simulation in the isolation BDT response; differences in the relative efficiency and \qq migration due to form factor uncertainties for both signal and normalisation channels; corrections to the \Lb kinematic properties; the disagreement in the \qq migration between data and simulation; and the finite size of the PID calibration samples. The total fractional systematic uncertainty is $\phantom{.}^{+7.8}_{-8.2}$\%, where the individual uncertainties are added in quadrature. The small impact of the form factor uncertainties means that the measured ratio of branching fractions can safely be considered independent of the theoretical input at the current level of precision.

From the ratio of yields and their determined efficiencies, the ratio of branching 
fractions of \PMuNu to \LcMuNu in the selected \qq regions is
\begin{equation}
\nonumber
\begin{split}
  \frac{\mathcal{B}(\PMuNu)_{\qq > 15\gevcc}}{\mathcal{B}(\LcMuNu)_{\qq > 7\gevcc}} = \phantom{,} \\
  (1.00\pm 0.04 \pm 0.08)\times10^{-2} \, , \phantom{}
\end{split}
\end{equation}
where the first uncertainty is statistical and the second is systematic. Using Eq.~\ref{eq:one} with $R_{\rm FF}=0.68\pm0.07$, computed in Ref.~\cite{LQCDdyn} for the restricted \qq regions, the measurement
\begin{displaymath}
\frac{\vub}{\vcb} = 0.083 \pm 0.004 \pm 0.004 \, ,
\end{displaymath}
is obtained. The first uncertainty arises from the experimental measurement and the second is due to the uncertainty in the LQCD prediction.  Finally, using the world average $\vcb = (39.5\pm0.8)\times 10^{-3}$ measured using exclusive decays~\cite{PDG2014}, \vub is measured as
\begin{displaymath}
\vub = (3.27\pm 0.15 \pm 0.16 \pm 0.06)\times 10^{-3} \, ,
\end{displaymath}
where the first uncertainty is due to the experimental measurement, the second arises from the uncertainty in the LQCD prediction and the third from the normalisation to \vcb. As the measurement of \vub/\vcb already depends on LQCD calculations of the form factors it makes sense to normalise to the \vcb exclusive world average and not include the inclusive \vcb measurements. The experimental uncertainty is dominated by systematic effects, most of which will be improved with additional data by a reduction of the statistical uncertainty of the control samples.

The measured ratio of branching fractions can be extrapolated to the full \qq region using \vcb and the form factor predictions~\cite{LQCDdyn}, resulting in a measurement of \mbox{$\mathcal{B}(\PMuNu) = (4.1\pm1.0)\times 10^{-4}$}, where the uncertainty is dominated by knowledge of the form factors at low \qq.

\begin{figure}[tb]
  \centering
  \includegraphics[width=0.45\textwidth]{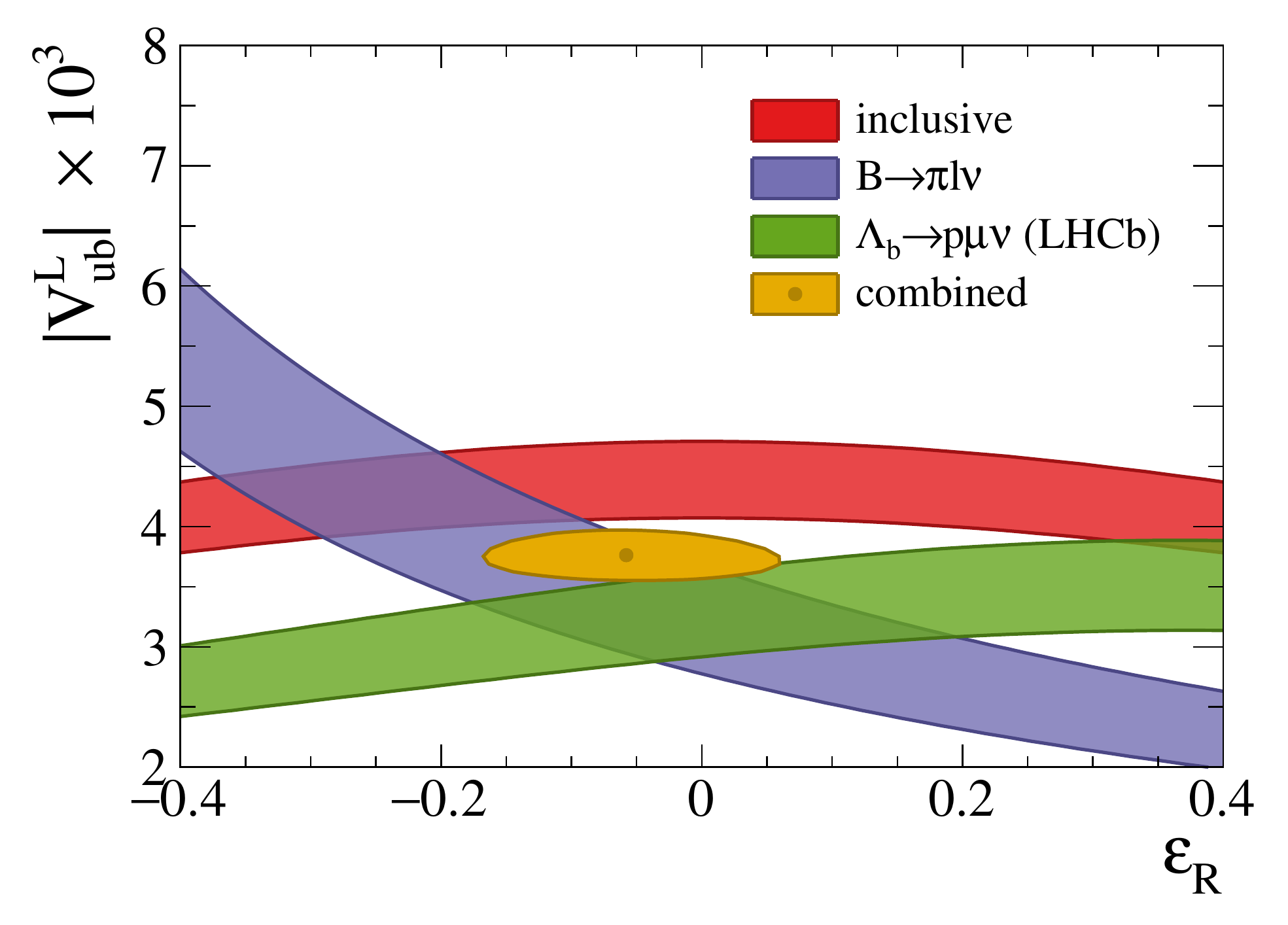} 
  \caption{\textbf{Experimental constraints on the left-handed coupling, \boldmath{$|V_{\uquark\bquark}^{\rm{L}}|$} and the fractional right-handed coupling, \boldmath{$\epsilon_R$}.} While the overlap of the 68\% confidence level bands for the inclusive~\cite{PDG2014} and exclusive~\cite{HFAG} world averages of past measurements suggested a right handed coupling of significant magnitude, the inclusion of the \lhcb \vub measurement does not support this.}
  \label{fig:righthanded}
\end{figure}

The determination of \vub from the measured ratio of branching fractions depends on the size of a possible right-handed coupling~\cite{Bernlochner:2014ova}. This can clearly be seen in Fig.~\ref{fig:righthanded}, which shows the experimental constraints on the left-handed coupling, $|V_{\uquark\bquark}^{\rm{L}}|$ and the fractional right-handed coupling added to the SM, $\epsilon_R$, for different measurements. The \lhcb result presented here is compared to the world averages of the inclusive and exclusive measurements. Unlike the case for the pion in $\Bzb\to\pip\ellm\neub$ and $\Bm\to\piz\ellm\neub$ decays, the spin of the proton is non-zero, allowing an axial-vector current, which gives a different sensitivity to $\epsilon_R$. The overlap of the bands from the previous measurements suggested a significant right-handed coupling but the inclusion of the \lhcb \vub measurement does not support that.

In summary, a measurement of the ratio of \vub to \vcb is performed using the exclusive decay modes \PMuNu and \LcMuNu. Using a previously measured value of \vcb, \vub is determined precisely. The \vub measurement is in agreement with the exclusively measured world average from Ref.~\cite{HFAG}, but disagrees with the inclusive measurement~\cite{PDG2014} at a significance level of 3.5 standard deviations. The measurement will have a significant impact on the global fits to the parameters of the CKM matrix.


\clearpage
\onecolumn

\section*{Acknowledgements}
\noindent
This article is dedicated to the memory of our dear friend and colleague, Till Moritz Karbach, who died following a climbing accident on 9th April 2015. Moritz contributed much to the physics analysis presented in this article. Within LHCb he was active in many areas; he convened the analysis group on beauty to open charm decays, he was deputy project leader for the LHCb Outer Tracker detector and he served the experiment as a shift leader. Moritz was a highly promising young physicist and we miss him greatly. We thank Stefan Meinel for a productive collaboration regarding form factor predictions of the \PMuNu and \LcMuNu decays, Winston Roberts for discussions regarding the \mbox{$\Lb\to N^{*}\mun\neumb$} decays and Florian Bernlochner for help in understanding the impact of right-handed currents. We express our gratitude to our colleagues in the CERN accelerator departments for the excellent performance of the LHC. We thank the technical and administrative staff at the LHCb institutes. We acknowledge support from CERN and from the national agencies: CAPES, CNPq, FAPERJ and FINEP (Brazil); NSFC (China); CNRS/IN2P3 (France); BMBF, DFG, HGF and MPG (Germany); INFN (Italy); FOM and NWO (The Netherlands); MNiSW and NCN (Poland); MEN/IFA (Romania); MinES and FANO (Russia); MinECo (Spain); SNSF and SER (Switzerland); NASU (Ukraine); STFC (United Kingdom); NSF (USA).  The Tier1 computing centres are supported by IN2P3 (France), KIT and BMBF (Germany), INFN (Italy), NWO and SURF (The Netherlands), PIC (Spain), GridPP (United Kingdom).  We are indebted to the communities behind the multiple open source software packages on which we depend. We are also thankful for the computing resources and the access to software R\&D tools provided by Yandex LLC (Russia).  Individual groups or members have received support from EPLANET, Marie Sk\l{}odowska-Curie Actions and ERC (European Union), Conseil g\'{e}n\'{e}ral de Haute-Savoie, Labex ENIGMASS and OCEVU, R\'{e}gion Auvergne (France), RFBR (Russia), XuntaGal and GENCAT (Spain), Royal Society and Royal Commission for the Exhibition of 1851 (United Kingdom).

\clearpage




\addcontentsline{toc}{section}{References}
\setboolean{inbibliography}{true}

\ifx\mcitethebibliography\mciteundefinedmacro
\PackageError{LHCb.bst}{mciteplus.sty has not been loaded}
{This bibstyle requires the use of the mciteplus package.}\fi
\providecommand{\href}[2]{#2}

\newpage
\centerline{\large\bf LHCb collaboration}
\begin{flushleft}
\small
R.~Aaij$^{38}$, 
B.~Adeva$^{37}$, 
M.~Adinolfi$^{46}$, 
A.~Affolder$^{52}$, 
Z.~Ajaltouni$^{5}$, 
S.~Akar$^{6}$, 
J.~Albrecht$^{9}$, 
F.~Alessio$^{38}$, 
M.~Alexander$^{51}$, 
S.~Ali$^{41}$
% ...

{\footnotesize \it
% ...
$ ^{5}$Clermont Universit\'{e}, Universit\'{e} Blaise Pascal, CNRS/IN2P3, LPC, Clermont-Ferrand, France\\
$ ^{6}$CPPM, Aix-Marseille Universit\'{e}, CNRS/IN2P3, Marseille, France\\
% ...
$ ^{9}$Fakult\"{a}t Physik, Technische Universit\"{a}t Dortmund, Dortmund, Germany\\
% ...
$ ^{37}$Universidad de Santiago de Compostela, Santiago de Compostela, Spain\\
$ ^{38}$European Organization for Nuclear Research (CERN), Geneva, Switzerland\\
% ...
$ ^{41}$Nikhef National Institute for Subatomic Physics, Amsterdam, The Netherlands\\
% ...
$ ^{46}$H.H. Wills Physics Laboratory, University of Bristol, Bristol, United Kingdom\\
% ...
$ ^{51}$School of Physics and Astronomy, University of Glasgow, Glasgow, United Kingdom\\
$ ^{52}$Oliver Lodge Laboratory, University of Liverpool, Liverpool, United Kingdom\\
% ...
}
\end{flushleft}

\end{document}